# Spin-orbit torques induced by interface-generated spin currents

Seung-heon C. Baek[1,2,*], Vivek P. Amin[3,4,*], Young-Wan Oh[1], Gyungchoon Go[5], Seung-Jae Lee[6], M. D. Stiles[4], Byong-Guk Park[1,**] & Kyung-Jin Lee[5,6,**]

[1]*Department of Materials Science and Engineering and KI for Nanocentury, KAIST, Daejeon 34141, Korea*

[2]*School of Electrical Engineering, KAIST, Daejeon 34141, Korea*

[3]*Maryland Nanocenter, University of Maryland, College Park, MD 20742, USA*

[4]*Center for Nanoscale Science and Technology, National Institute of Standards and Technology, Gaithersburg, Maryland 20899, USA*

[5]*Department of Materials Science and Engineering, Korea University, Seoul 02841, Korea*

[6]*KU-KIST Graduate School of Converging Science and Technology, Korea University, Seoul 02841, Korea*

*These authors equally contributed to this work.

**Corresponding emails: bgpark@kaist.ac.kr (B.-G. Park) and kj_lee@korea.ac.kr (K.-J. Lee)

**Magnetic torques generated through spin-orbit coupling[1-8] promise energy-efficient spintronic devices. It is important for applications to control these torques so that they switch films with perpendicular magnetizations without an external magnetic field[9-14]. One suggested approach[15] uses magnetic trilayers in which the torque on the top magnetic layer can be manipulated by changing the magnetization of the bottom layer. Spin currents generated in the bottom magnetic layer or its interfaces transit the spacer layer and exert a torque on the top magnetization. Here we demonstrate field-free switching in such structures and attribute it to a novel spin current[16,17] generated at the interface between the bottom layer and the spacer layer. The measured torque has a distinct dependence on the bottom layer magnetization which is consistent with this interface-generated spin current but not the anticipated bulk effects[15]. This other interface-generated spin-orbit torque will enable energy-efficient control of spintronic devices.**

Spin current generation by the spin-orbit interaction is a central theme in condensed matter physics[18]. Two fundamental questions about spin current generation via the spin-orbit interaction relate to modifying the spin polarization carried by the spin current. First, how can one increase the magnitude of spin polarization, i.e., the conversion efficiency from a charge current to a spin current? Most studies have focused on this objective, which typically involves searching for materials with large spin Hall effect[1-8], which converts a charge current to a spin current[19,20] through bulk spin-orbit coupling. In this paper, we address a second question: How can we control the direction of the spin polarization?

Current implementations of magnetic high density memory and logic applications use structures with magnetizations perpendicular to the film[21-23]. For commercial viability, it is necessary to switch the state of such structures without applying an external magnetic field. Deterministic field-free switching of perpendicular magnetizations via spin-orbit torques is impossible unless an in-plane magnetic field[1] (or effective field[9-14]) is applied or, as we demonstrate, the spin σ of the incoming spin current has a component anti-aligned with the perpendicular magnetization. In isotropic materials, symmetry requires that for the spin Hall effect, the spin polarization σ, spin-current flow, and charge-current flow are mutually orthogonal. For example, charge flowing in the electric field direction (*x*-direction) generates spin flowing toward the interface normal (*z*-direction) and this spin current is spin-polarized along the σ = ±**y** direction. While a recent report demonstrated experimentally that σ of the spin currents generated in WTe$_2$ deviates from **y** due to the crystal symmetry[24], engineering the spin polarization direction has been largely unexplored for sputtered metallic heterostructures that lack a well-defined crystal orientation and that are of technological importance.

In this work, we demonstrate theoretically that ferromagnet (FM)/normal metal interfaces *can* generate a spin current in the normal metal that has an out-of-plane (*z*) component of the spin polarization in addition to an in-plane (*y*) component, and demonstrate experimentally that they *do*. This effect (Supplementary Note 1) arises from a combination of two processes[16,17]. First, the in-plane electric field (**E**//**x**) creates non-equilibrium carriers that are anisotropic in momentum space and differ between the ferromagnetic and normal metal layers (because of their different electrical conductivities). The asymmetry between carriers in different layers allows for net propagation normal to the interface, perpendicular to the electric field. Second, carriers scattering off the interface interact with interfacial spin-orbit fields, polarizing the flow of spins. These processes enable an in-plane electric field to generate a spin current flowing out-of-plane.

Two distinct mechanisms are important for electron spins scattering from an interface: spin-orbit filtering and spin-orbit precession. The former applies to the component of the spins along the interfacial spin-orbit field and the latter to the transverse components. Carriers incident to the interface with spins parallel and antiparallel to the field have different reflection and transmission probabilities. After scattering, an unpolarized current becomes polarized. When summed over all electrons, this spin-orbit filtering gives a net spin polarization in the $\mathbf{y} = \mathbf{z} \times \mathbf{E}$ direction, identical to that of the spin Hall spin current (Supplementary Note 1).

Spin-orbit precession occurs because incoming carriers with opposite spins perpendicular to the spin-orbit field both precess the same while scattering off the interface. If the incoming current has no net spin polarization, no polarization develops. However, if the incoming current has a net polarization, such as from the ferromagnetic layer of ferromagnet/normal metal bilayers, then after precession, the net polarization survives and changes its orientation. After summing over the Fermi surfaces, the spin-orbit precession current has a net spin polarization in the $\mathbf{m} \times \mathbf{y}$ direction where $\mathbf{m}$ is the magnetization vector of ferromagnetic layer (Supplementary Note 1). For an in-plane magnetized ferromagnet ($\mathbf{m}//\mathbf{x}$), this mechanism generates a spin current flowing into the normal metal polarized with a $z$-component.

Symmetry does allow for similar spin currents in bulk ferromagnets. However, the spin currents generated by the bulk spin-orbit interaction are expected to have spins largely aligned with the magnetization[15] because precession of the spins around the exchange field rapidly dephases the transverse components. Interfacially generated spin currents are not subject to dephasing in the normal metal, allowing for much larger components transverse to the magnetization.

To test the prediction of spin currents generated at the interface, we measure spin-orbit torques for bottom FM (4 nm)/Ti (3 nm)/top CoFeB (1 nm to 1.4 nm)/MgO (1.6 nm) Hall bar structures. The top CoFeB layer has perpendicular magnetic anisotropy and serves as a spin current analyzer while the bottom FM is an in-plane magnetized CoFeB or NiFe layer (Fig. 1a and Methods). We refer to these structures collectively as FM/Ti samples and particularly as CoFeB/Ti samples or NiFe/Ti samples. We choose these structures because the insertion of a Ti layer adds an additional FM/Ti interface but as we show below, the Ti layer itself generates a negligible spin current. Consequently, any spin current generated in the FM/Ti samples is caused either by the bulk spin-orbit interaction of the bottom ferromagnet[15] or by the interfacial spin-orbit interaction of the FM/Ti interface[16,17].

We first perform harmonic Hall voltage measurements[4,5] (Methods) to assess the damping-like and field-like spin-orbit torques. We also measure spin-orbit torque switching as an independent test for the sign of the spin-orbit torque. In the harmonic Hall measurement with an ac current applied in the *x*-direction, the sign of the 2$^{nd}$ harmonic signal ($V_{2\omega}$) for an in-plane magnetic field $B = B_x$ ($B = B_y$) gives the sign of damping-like (field-like) spin-orbit torque (see schematic in Fig. 1a). We examine four types of samples: the CoFeB/Ti sample, the NiFe/Ti sample, and two other types of samples, in which the FM/Ti bilayer is replaced by a single Ta or Ti layer (i.e., the Ta sample and the Ti sample). The Ta sample provides a reference sample for the sign of spin-orbit torque.

The Ta sample shows a negative peak in the 2$^{nd}$ harmonic signal for a positive in-plane field (i.e., $B_x > 0$), corresponding to a negative spin Hall angle (Fig. 1b). Spin-orbit torque switching of the Ta sample shows up-to-down switching for negative current in the presence of positive $B_x$ (Fig. 1f), which is applied to give deterministic switching[1]. This switching direction also corresponds to a negative spin Hall angle (Fig. 1b). On the other hand, the Ti sample shows negligible spin current generation as seen both in the lack of spin-orbit torque switching (Figs. 1c) and in the small 2$^{nd}$ harmonic signal $V_{2\omega}$ when normalized by the maximal change in $V_{1\omega}$ shown in the insets[4,5].

Importantly, we find that the CoFeB/Ti sample (Figs. 1d and 1h) and the NiFe/Ti sample

(Fig. 1e and 1i) exhibit spin-orbit torques sufficiently large to switch the perpendicular magnetization of the top CoFeB layer. The anomalous Nernst contribution to the 2$^{nd}$ harmonic voltage, induced by the bottom in-plane FM layer, has been removed in Figs. 1d and 1e (Supplementary Note 2). A difference between the CoFeB/Ti and NiFe/Ti samples is the sign of spin-orbit torque, i.e., the sign of spin polarization. The CoFeB/Ti sample shows the same sign as the Ta sample but the NiFe/Ti sample has the opposite sign. This sign difference between the samples with nominally identical structures except for the type of ferromagnet unambiguously demonstrates that the spin current generated from the bulk ferromagnet or FM/Ti interface is responsible for the spin-orbit torque. We estimate the effective spin Hall angles (Supplementary Note 3) as ≈ −0.048±0.002 for the Ta sample, ≈ −0.014±0.001 for the CoFeB/Ti sample, and ≈ +0.006±0.0006 for the NiFe/Ti sample (uncertainties are single standard deviations). Therefore, the effective spin Hall angles of the FM/Ti samples are non-negligible.

We next examine the possibility that the spin current in FM/Ti samples originates from a bulk spin-orbit interaction of the bottom ferromagnet. Comparing in-plane magnetized CoFeB and NiFe layers without a perpendicularly magnetized top CoFeB layer, we find that the anomalous Hall signals are of the opposite sign, consistent with a previous calculation[25], whereas the anisotropic magnetoresistance signals are of the same sign (Supplementary Note 4). Given that the spin-orbit torque sign is different between the CoFeB/Ti and NiFe/Ti samples (Fig. 1), the spin current must originate from the anomalous Hall effect if it is generated by the bulk spin-orbit interaction of bottom ferromagnetic layer.

The spin polarization direction of the spin current originating from the anomalous Hall effect is expected to align with the magnetization direction of the ferromagnet[15] and can be analyzed through the 2$^{nd}$ harmonic signal as a function of the azimuthal angle of magnetization in the ferromagnet. Macrospin modeling gives the expected variation of 2$^{nd}$ harmonic signals with the azimuthal angle of magnetization for a fixed spin direction ($\sigma$ = **y**; Fig. 2a) and for the anomalous Hall effect (Fig. 2b). Figures 2c and 2d show the measured 2$^{nd}$ harmonic signals as a function of the azimuthal angle of in-plane magnetic field for the Ta sample and the CoFeB/Ti sample, respectively. We find that the samples behave similarly to the calculation for the fixed $\sigma$ = **y**. The NiFe/Ti sample exhibits similar dependence but with reversed sign (Supplementary Note 5). From these results, we conclude that the spin current in the FM/Ti samples appears to have its spin component along the **y** direction, which is not consistent with the expected behavior of the anomalous Hall effect, but is consistent with what we expect from the interfacial spin-orbit interaction of a FM/Ti interface.

A remaining task is to examine if the spin polarization of interface-generated spin current has an additional *z*-component ($\sigma_z$) as predicted by theory. To test this, we measure a hysteresis loop of the anomalous Hall signal $R_{xy}$ (i.e., $m_z$ component of the top perpendicular CoFeB layer) versus an out-of-plane field $B_z$ in the presence of a d.c. current. We note that a spin current flowing out-of-plane with a spin *z*-component serves as an anti-damping torque for a perpendicular magnetization and one current polarity. Anti-damping causes an abrupt increase in the loop shift as a function of dc current $I_{dc}$ at a threshold above which it exceeds the intrinsic damping, as in conventional spin-transfer torque studies[26] and is also indicated by down arrows in modeling results (Fig. 3a). Here we define the center of the hysteresis loop $B_S(I_{dc}) = \left[\left|B_C^+(I_{dc})\right| - \left|B_C^-(I_{dc})\right|\right]/2$ where $B_C^{\pm}$ are positive and negative magnetization reversal fields, and the loop shift $\Delta B_S(I_{dc}) = B_S(I_{dc}^+) - B_S(I_{dc}^-)$ where $I_{dc}^{\pm}$ are positive and negative dc currents. We note that such a threshold effect is absent for $\sigma$ = **y** and external in-plane field $B_x$ = 0 (Black solid square symbols in Fig. 3a). We also note that for the case with $\sigma$ = **y** and $B_x$

≠ 0, $\Delta B_s$ gradually increases with dc current but there is no threshold effect (Black open circular symbols in Fig. 3a). In Figs. 3b and 3c, we show that the threshold effect is observed experimentally for the CoFeB/Ti sample. The hysteresis loops remain the same for dc currents up to 5 mA and then start to shift to the positive (negative) $B_z$ direction for a larger positive (negative) dc currents when the magnetization direction of the in-plane CoFeB is set in the positive *x* direction. The direction of the loop shift reverses when changing the magnetization direction of the in-plane CoFeB to the negative *x* direction, which is consistent with the theoretical prediction, i.e., $\sigma_z \sim \mathbf{m} \times \mathbf{y}$. This threshold effect differs from the linear dependence of $\Delta B_S$ on the dc current for the Ta sample in the presence of $B_x$ (Black open circular symbols in Fig. 3c and Supplementary Note 6).

Spin-orbit torque switching without in-plane magnetic fields provides additional support for this spin *z*-component. As the spin *z*-component favors opposite magnetization directions of the top CoFeB layer for opposite current directions, it enables field-free spin-orbit torque switching. In Figs. 3d and 3e, we show that field-free switching is achieved for the CoFeB/Ti sample when the magnetization of the bottom, in-plane layer points along the +*x* and -*x* directions, respectively. We note that stray fields from the in-plane CoFeB layer could cause field-free switching but must show a linear increase with dc current even below 5 mA, which is not seen in Fig. 3c. The threshold effect in $\Delta B_S$ together with field-free switching proves the existence of a spin *z*-component in the polarization of the interface-generated spin currents.

In this work, we demonstrate the interface-generated spin current. We expect the previously observed inverse spin Hall effect in ferromagnets[27] is related to the interface-generated spin current. Moreover, as widely-studied ferromagnet/heavy metal bilayers also have an interface, we expect that a non-negligible interface-generated spin current is present in bilayers as well, as recently suggested by *ab-initio* studies[28,29]. Our finding of the interface-generated spin current will broaden the scope of material engineering for spintronic devices, and be beneficial for spin-orbit torque switching devices with perpendicular magnetization by eliminating the external field that would be deleterious to high-density device integration.

**Note**

While we were preparing the manuscript, we became aware of similar work done by another group[30].

**Methods**

**Sample preparation** The samples of underlayer/FM(4nm)/Ti(3nm)/CoFeB/MgO, Ti/CoFeB/MgO, Ta/CoFeB/MgO, and Ti/CoFeB/MgO structures were prepared on thermally oxidized Si substrates by magnetron sputtering with a base pressure of less than $4.0 \times 10^{-6}$ Pa ($3.0 \times 10^{-8}$ Torr) at room temperature. A underlayer of Ti (2 nm to 4 nm) was introduced for FM/Ti samples to improve the adhesion of FM layer on $SiO_2$ substrate and a capping layer of Ta (2 nm) was used to protect the MgO layer. All metallic layers were grown by d.c. sputtering with a working pressure of 0.4 Pa (3 mTorr), while the MgO layer is deposited by RF sputtering (150 W) from an MgO target at 1.33 Pa (10 mTorr). The compositions of CoFeB and NiFe are $Co_{32}Fe_{48}B_{20}$, and $Ni_{81}Fe_{19}$, respectively. All samples were annealed at 150 °C for 40 min in vacuum condition, $4.0 \times 10^{-4}$ Pa ($3.0 \times 10^{-6}$ Torr), to promote the perpendicular magnetic anisotropy. The Hall-bar structured devices including a square-shaped ferromagnetic island were fabricated using photo-lithography and Ar ion-beam etching. The width of the Hall bar is 5 μm and the size of the ferromagnetic island is $4 \times 4$ μm$^2$.

**Spin-orbit torque measurements** The spin-orbit torque was characterized using a harmonic lock-in technique. The first and second harmonic Hall resistances for an a.c. current of 50 Hz were simultaneously measured while sweeping the in-plane external magnetic field, in the longitudinal ($B_x$) or transverse ($B_y$) direction to the current direction. The in-plane magnetic field has a slight out-of-plane tilt angle (2° to 4°) from the film plane, which prevents multidomain formation. The single standard deviation uncertainty of the lock-in harmonic Hall voltage measurements is ±0.15 µV. Corresponding error bars are included in the figures. In most cases, the error bars are smaller than symbols in the figures. The SOT-induced switching experiments were done by measuring the anomalous Hall resistance using a d.c. current of 100 µA after applying a current pulse of 20 µs with a fixed $B_x$. All measurements were carried out at room temperature. More than three samples are measured for each type of sample; data are qualitatively reproducible.

**Numerical Simulations** For Figs. 2a and 2b (harmonic Hall signals), we carried out macrospin simulations by solving the Landau-Lifshitz-Gilbert (LLG) equation in the presence of an external magnetic field and a spin-transfer torque from the spin Hall effect (Fig. 2a) or the anomalous Hall effect of the FM layer (Fig. 2b). For the spin-transfer torques due to the spin Hall effect, we considered both damping-like torques (DLT) and field-like torques (FLT) ($FLT/DLT = -3.5$). For the spin-transfer torques due to the anomalous Hall effect, we adopted the theory of Ref. [15]. We used the following parameters for CoFeB: the saturation magnetization $M_s$ = 800 kA m$^{-1}$, the perpendicular anisotropy field $\mu_0 H_K$, = 1.15 T, the anomalous Hall conductivity $\sigma_{AH}/\sigma = -0.001$, the spin polarization of longitudinal transport $\beta = 0.56$ and the anomalous Hall effect $\zeta = 0.7$, the spin mixing conductances Re[$G^{\uparrow\downarrow}$] =

$3.9\times10^{14}\,\Omega^{-1}\mathrm{m}^{-2}$, $\mathrm{Im}[G^{\uparrow\downarrow}]=0.39\times10^{14}\,\Omega^{-1}\mathrm{m}^{-2}$, and the spin diffusion length $l_{sd}^{F}=5.5\,\mathrm{nm}$. The in-plane external magnetic field has an out-of-plane tilt angle of 3° from the film plane.

For Fig. 3a (loop-shift field $\Delta B_S$ versus dc current), we numerically solved the LLG equation including a spin torque [$\sim \mathbf{m}\times(\mathbf{m}\times\boldsymbol{\sigma})$] for a semi-one dimensional system that is discretized only along the current direction. We used the following parameters for the simulations: $M_s$ = 1000 kA m$^{-1}$, the exchange stiffness constant $A_{\mathrm{ex}} = 1.6\times10^{-11}$ J m$^{-1}$, the Gilbert damping constant = 0.05, the effective spin Hall angle = -0.014, the perpendicular anisotropy $K_U = 1\times10^6$ J m$^{-3}$, the unit cell size = 4 nm × 400 nm × 1.2 nm, and the number of cells along the current direction = 100.

**Acknowledgements**

The authors acknowledge K.-W. Kim, H.-W. Lee, and J. Sinova for discussion. This work was supported by Creative Materials Discovery Program through the National Research Foundation of Korea (NRF-2015M3D1A1070465). B.-G.P. acknowledges financial support from the NRF (NRF-2014R1A2A1A11051344), and K.-J.L. from the NRF (NRF-2017R1A2B2006119). V.A.

acknowledges support under the Cooperative Research Agreement between the University of Maryland and the National Institute of Standards and Technology Center for Nanoscale Science and Technology, Award 70NANB14H209, through the University of Maryland.

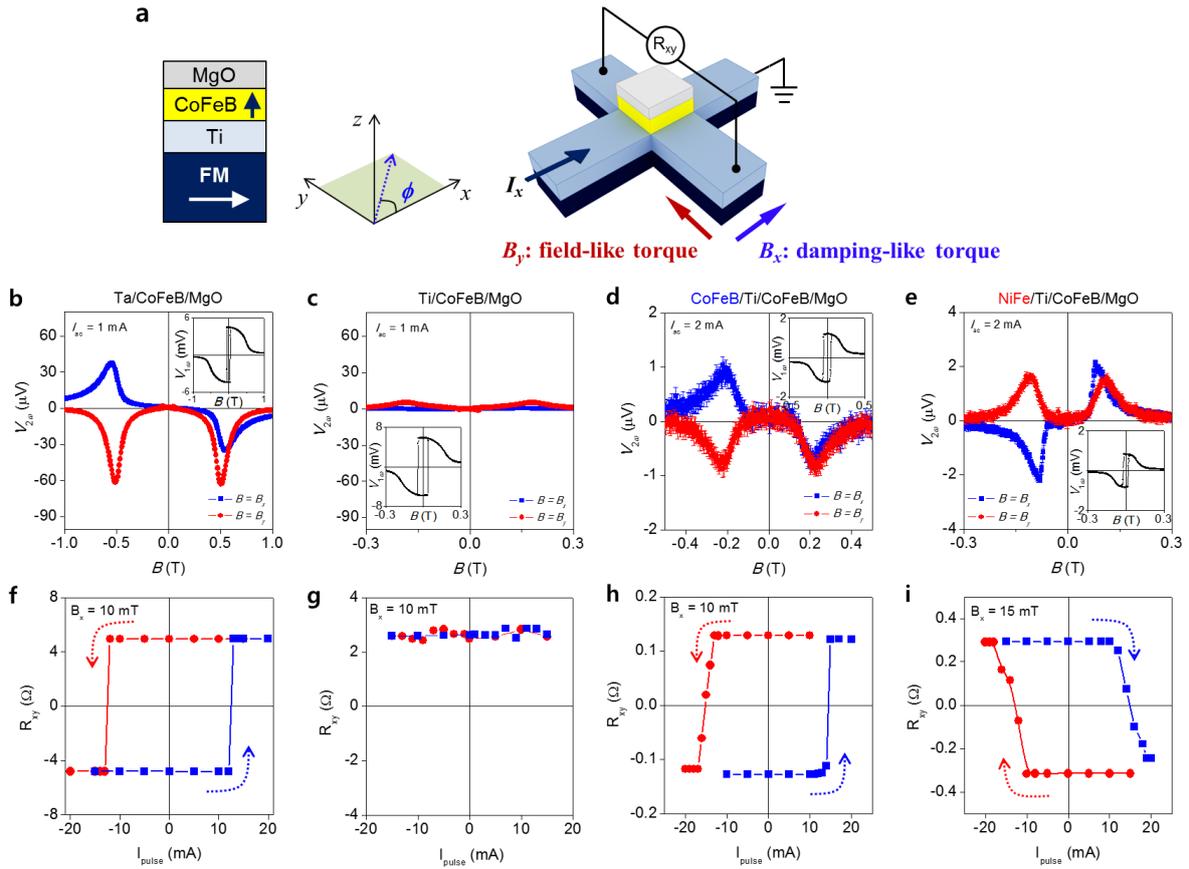

**Figure 1| Spin-orbit torques in ferromagnet (FM)/Ti/CoFeB/MgO samples (FM = CoFeB or NiFe). a**, Schematics of the FM/Ti/CoFeB/MgO layer (left) and spin-orbit torque measurement in Hall bar structure (right). $I_x$ is the in-plane current and $\phi$ is the azimuthal angle. $\phi = 0°$ (90°) for the in-plane magnetic field $B_x$ ($B_y$). The 2$^{nd}$ harmonic signal $V_{2\omega}$ for **b,** the Ta(3 nm)/CoFeB/MgO, **c,** Ti(3 nm)/CoFeB/MgO, **d,** CoFeB(4 nm)/Ti(3 nm)/CoFeB/MgO, and **e,** NiFe(4 nm)/Ti(3 nm)/CoFeB/MgO samples. The insets show the 1$^{st}$ harmonic signals $V_{1\omega}$ with an ac current $I_{ac}$. The switching experiment under $B_x$ for **f,** the Ta(3 nm)/CoFeB/MgO, **g,** Ti(3 nm)/CoFeB/MgO, **h,** CoFeB(4 nm)/Ti(3 nm)/CoFeB/MgO, and **i,** NiFe(4 nm)/Ti(3 nm)/CoFeB/MgO samples. The magnetization direction of the top CoFeB layer is monitored by measuring the anomalous Hall resistance $R_{xy}$ while sweeping a pulsed current $I_{pulse}$. Blue and red dotted arrows indicate the switching direction. Error bars, many smaller than the

symbols, indicate single standard deviation uncertainties.

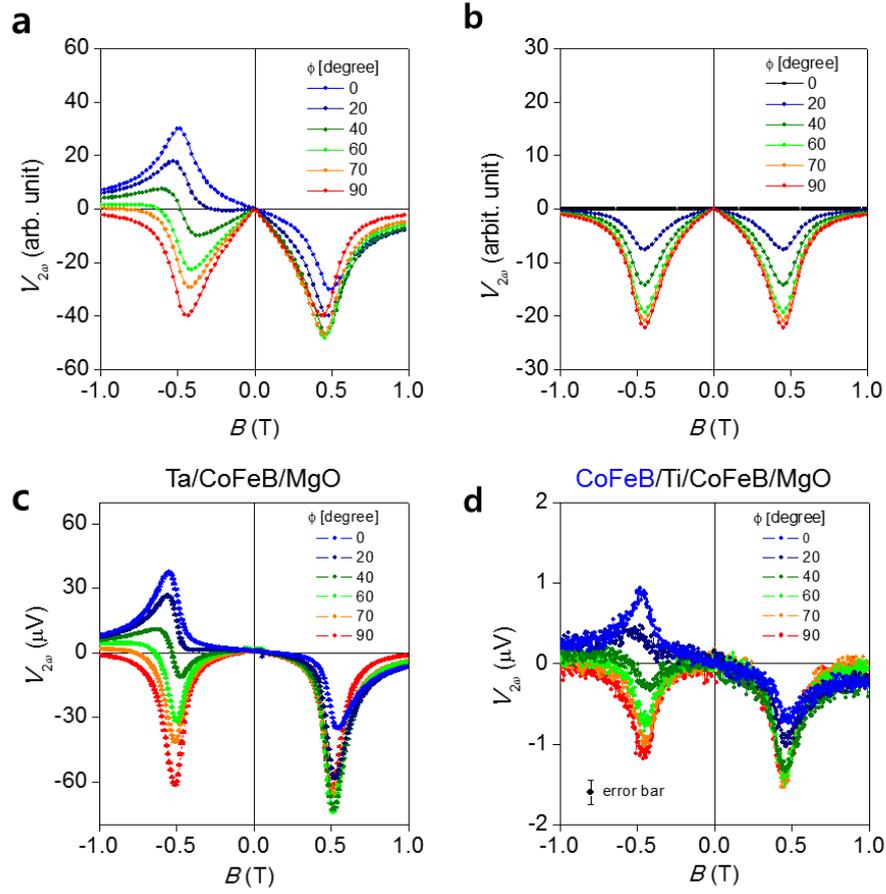

**Figure 2 | Azimuthal angle-dependent $V_{2\omega}$ in the CoFeB(4 nm)/Ti (3 nm)/CoFeB/MgO-sample.** Macrospin modelling results of $V_{2\omega}$ for **a,** the bulk spin Hall effect ($\sigma = \mathbf{y}$) and **b,** the anomalous Hall effect of bulk FM layer, as a function of the azimuthal angle $\phi$. Experimentally measured results of $V_{2\omega}$ for **c,** the Ta(3 nm)/CoFeB/MgO sample and **d,** CoFeB(4 nm)/Ti(3 nm)/CoFeB/MgO sample, as a function of the azimuthal angle $\phi$.

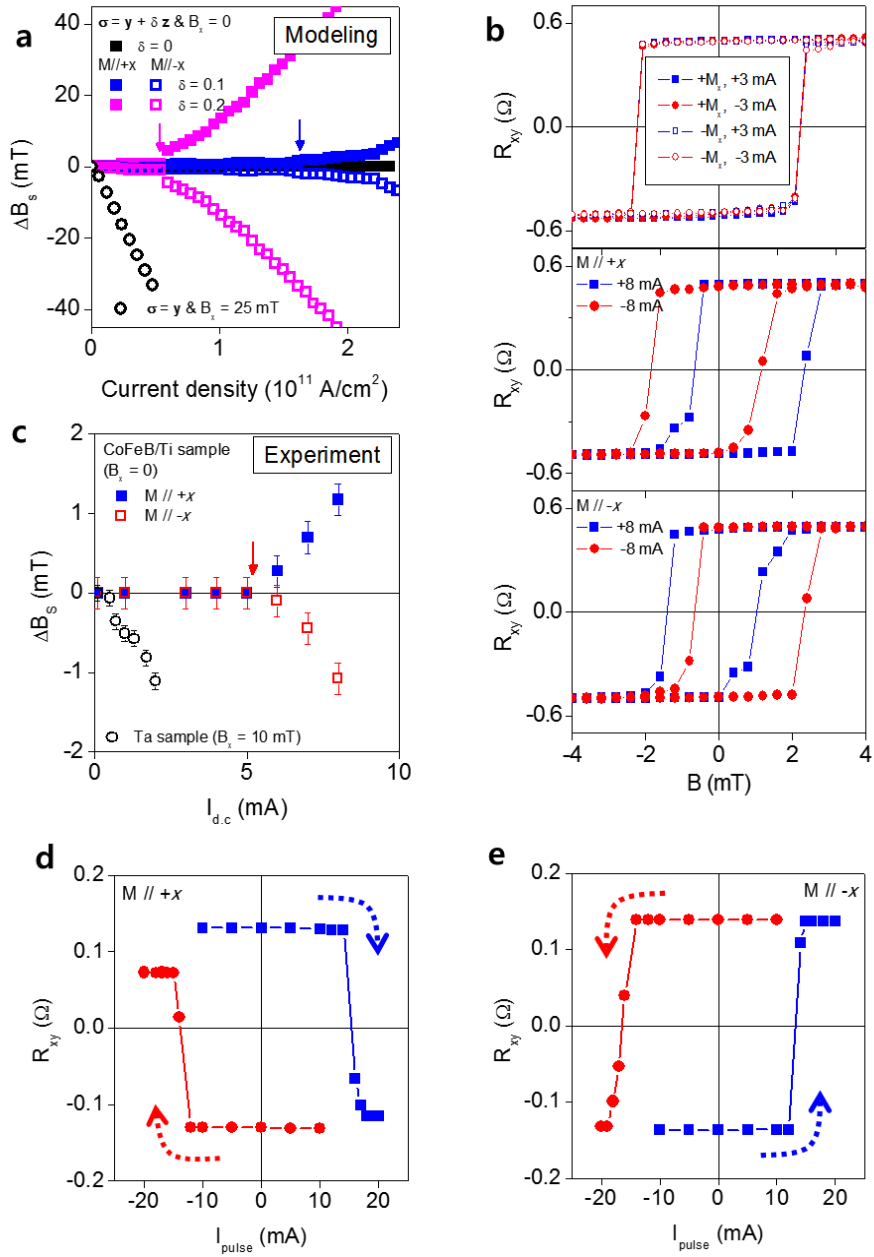

**Figure 3 | The spin-z component of interface-generated spin currents. a,** macrospin simulation results of the loop-shift field $\Delta B_S$ versus d.c current density for cases of the interface-generated spin current ($\sigma = \mathbf{y} + \delta\,\mathbf{z}$ where $\delta$ is the ratio of the spin-$z$ component to the spin-$y$ component; square symbols) and of the bulk spin Hall effect ($\sigma = \mathbf{y}$; open circular symbols). The loop-shift field $\Delta B_S$ is defined as the difference in the centres of the hysteresis

loop for an in-plane d.c. current $+I_{dc}$ and $-I_{dc}$. We note that $B_x$ is zero (20 mT) for case of the interface-generated spin current (bulk spin Hall effect). **b,** Experimental measurements of $R_{xy}$ versus $B_z$ curves: (top panel) $I_{dc}$ of ±3 mA, (middle panel), $I_{dc}$ of ±8 mA and magnetization of the bottom CoFeB layer (*M*) // +x direction, and (bottom panel) $I_{dc}$ of ±8 mA and *M* // −x. **c,** Experimental $\Delta B_S$ versus $I_{dc}$. Blue (red) square symbols represent the results for *M* // +x (−x) of the CoFeB/Ti sample when $B_x = 0$. Black open circular symbols are of the Ta sample under $B_x$ =10 mT. Down arrows in **a** and **c** represent threshold d.c. currents above which $\Delta B_S$ abruptly changes. Experimental spin-orbit torque switching in the CoFeB(4 nm)/Ti(3 nm)/CoFeB/MgO sample without an external magnetic field for **d**, *M* // +x and **e**, *M* // −x. Error bars indicate single standard deviation uncertainties.

# Supplementary Information

− Contents −

**Note 1. Theoretical background of interface-generated spin currents**

In heavy metal/ferromagnetic bilayers, spin-orbit torques are typically separated into two categories: those that arise from the spin Hall effect [S1, S2] and those that arise from the Rashba-Edelstein effect [S3, S4]. In the presence of an in-plane electric field, the spin Hall effect generates a spin current in the heavy metal that flows out-of-plane and exerts a spin transfer torque on the ferromagnetic layer. In the same geometry, the Rashba-Edelstein effect generates a spin accumulation carried by a two-dimensional electron gas trapped at the interface; this spin accumulation exerts a torque directly on the ferromagnetic layer via the exchange interaction. In this work, we experimentally demonstrate a third possibility, in which the interface between the heavy metal and the ferromagnet generates a spin current through a process physically distinct from the spin Hall or Rashba-Edelstein effects.

The most important characteristic of this interface-generated spin current is that it exerts spin-orbit torques not bound by the same symmetry constraints as the other known mechanisms. While this spin current exerts a spin torque on the ferromagnetic layer of a heavy metal/ferromagnetic bilayer, it is difficult to experimentally distinguish this torque from the other torques caused by the spin Hall and Rashba-Edelstein mechanisms. To circumvent this difficulty, we investigate torques in ferromagnet (FM1)/nonmagnet/ferromagnet (FM2) spin valves driven by an in-plane electric field. In this scheme, the interface between a fixed ferromagnetic layer (FM1) and the nonmagnet generates a spin current while the other ferromagnetic layer (FM2) receives the resulting spin torque.

The interface-generated spin current arises from a combination of two processes [S5, S6]. First, the in-plane electric field creates a non-equilibrium occupation of carriers that is anisotropic in carrier momentum. Second, as carriers scatter off the interface, they undergo momentum-dependent spin filtering and momentum-dependent spin precession while interacting with the interfacial spin-orbit field. The combination of these two processes (anisotropic occupation and spin-orbit scattering) results in a net spin current.

Spin-orbit filtering currents occur because carriers with spins that are parallel or antiparallel to the interfacial spin-orbit field have different reflection and transmission probabilities. Thus, an incoming current of unpolarized carriers becomes spin polarized upon reflection and transmission. This process is easiest to understand in nonmagnetic bilayers, in which an arbitrary quantization axis can be chosen for each incoming state. However, the effect persists even if the incoming states are spin-split, as is the case if one of the layers is ferromagnetic. After summing over the relevant states, the scattered carriers have a net spin polarization along the $\boldsymbol{f} = \boldsymbol{z} \times \boldsymbol{E}$ direction, where $\boldsymbol{z}$ is the interface normal. This polarization direction is identical to that of the spin Hall current and the spin accumulation caused by the Rashba-Edelstein effect.

Spin-orbit precession currents occur because carriers precess about the axes aligned with the spin-orbit fields while scattering off the interface. In this case, incoming carriers will change their spin orientation upon scattering, but if the incoming current is unpolarized then the scattered current also remains unpolarized. However, if the incoming current from at least one of the layers is spin polarized, then the reflected and transmitted carriers change their spin orientation and remain spin polarized upon scattering. Thus, the spin-orbit precession current only occurs if one of the two layers is ferromagnetic. The spin-orbit precession current is proportional to the polarization (P) of the ferromagnetic layer and has a net spin polarization aligned along the $\boldsymbol{m} \times \boldsymbol{f}$ direction.

The total interface-generated spin current results from a combination of spin-orbit filtering and spin-orbit precession and has the following form:

$$\boldsymbol{j} = j_f \boldsymbol{f} + j_p P \boldsymbol{m} \times \boldsymbol{f}, \qquad (S1)$$

where $j_f$ and $j_p$ give the strengths of the spin-orbit filtering current and spin-orbit precession current, respectively. The spin current is expressed as a vector which points along the direction of spin polarization, and the flow direction is assumed to be out of plane (*z*). The magnitudes of both $j_f$ and $j_p$ are magnetization-independent in the model introduced in Refs. [S5, S6] when the interfacial exchange interaction vanishes, but can be magnetization-dependent for more complicated models.

Spin currents that have out-of-plane spin polarizations are highly desirable for efficiently switching perpendicularly-magnetized ferromagnetic layers. As can be seen from Eqn. (S1), the spin-orbit precession current carries an out-of-plane spin polarization when the magnetization has an in-plane component. For example, if the magnetization and the electric field both point along *x*, then the spin polarization of the spin-orbit precession current points along $\boldsymbol{m} \times (\boldsymbol{z} \times \boldsymbol{E}) = -\boldsymbol{z}$. The strength and sign of this spin current are determined by details of the electronic structures of each layer and by interfacial properties [S5, S6]. The anomalous Hall effect can generate a spin current that flows out-of-plane and has an out-of-plane spin polarization, but only if the magnetization also has an out-of-plane component [S7]. In contrast, the spin-orbit precession current naturally has the desired orientation at interfaces between nonmagnets and ferromagnets with in-plane anisotropy. To incite switching of perpendicular layers thus requires a FM1/NM/FM2 trilayer, in which the first ferromagnetic layer (FM1) is in-plane and fixed while the second ferromagnetic layer (FM2) is out-of-plane and free to switch.

To derive the interface-generated spin current, we use the formalism developed in [S6]. In that paper, both the nonmagnet and ferromagnet are modeled as spin-polarized free electron gases with identical, spin-independent, spherical Fermi surfaces. The nonequilibrium occupation of carriers incident to the interface is polarized in the ferromagnet and unpolarized in the nonmagnet. We treat the scattering potential as a delta function in $z$ that has the following form:

$$V(\boldsymbol{r}) = \frac{\hbar^2 k_F}{m} \delta(z)(u_0 + u_R \boldsymbol{\sigma} \cdot (\widehat{\boldsymbol{k}} \times \widehat{\boldsymbol{z}})). \tag{S2}$$

Here $u_0$ is the strength of a spin-independent barrier, $u_R$ is the scaled Rashba parameter, $\boldsymbol{\sigma}$ is the Pauli vector, $\widehat{\boldsymbol{k}}$ is a unit vector pointing along the incident momentum, and $\delta(z)$ is the delta function. Note that in comparison to the scattering potential used in [S6], we have removed the interfacial exchange interaction $u_{ex}$ because doing so greatly simplifies the calculation. Although adding an interfacial exchange interaction and making the Fermi surfaces in the ferromagnet spin-dependent does change the form of the interface-generated spin current, it does not qualitatively alter the result needed for the experimental analysis of this paper.

We begin with the expression for the spin current just within the nonmagnetic metal ($z = 0^-$), as given by Eqn. (B20) in [S6]:

$$j_\sigma = \frac{e}{\hbar}\left(\frac{v_F}{2\pi}\right)^3 E \int d\bar{k}_x d\bar{k}_y \, \bar{k}_x [\tau^{FM} PT(\boldsymbol{k})_{\sigma\sigma'} m_{\sigma'} + (\tau^{NM} - \tau^{FM}) T(\boldsymbol{k})_{\sigma c}]. \tag{S3}$$

Note that the equation produced here is equivalent to Eqn. (B20) in [S6] but is rewritten for the purposes of this calculation. Here $E$ is the electric field (assumed to point along the $x$-axis), $v_F$ is the Fermi velocity, $\tau^{NM/FM}$ is the momentum relaxation time of the nonmagnet/ferromagnet, $P$ is the polarization of the ferromagnet, and $m_{\sigma'}$ are the components of the magnetization of the ferromagnet. The notation $\bar{k}_i \equiv k_i/k_F$ applies for $i \in [x, y, z]$, where $k_F$ is the Fermi momentum. The subscript $\sigma$ runs along the three components of spin polarization, and can be treated in any reference frame that is convenient. The tensors $T_{\sigma c}$ and $T_{\sigma\sigma'}$ are functions of the reflection and transmission coefficients at the interface, defined as follows

$$T(\boldsymbol{k})_{\sigma c} = \frac{1}{2}\mathrm{tr}[t(\boldsymbol{k})^\dagger \sigma_\sigma t(\boldsymbol{k})], \tag{S4}$$

$$T(\boldsymbol{k})_{\sigma\sigma'} = \frac{1}{2}\mathrm{tr}[t(\boldsymbol{k})^\dagger \sigma_\sigma t(\boldsymbol{k}) \sigma_{\sigma'}]. \tag{S5}$$

The matrices $t$ are the 2×2 $k$-dependent transmission matrices, which relate the incoming spinor to the outgoing spinor at each k point,

$$t(\boldsymbol{k}) = \begin{pmatrix} t^\uparrow(\boldsymbol{k}) & 0 \\ 0 & t^\downarrow(\boldsymbol{k}) \end{pmatrix}, \tag{S6}$$

$$t^{\uparrow/\downarrow}(\boldsymbol{k}) = \frac{i\bar{k}_z}{i\bar{k}_z - (u_0 \pm u_{eff}(\boldsymbol{k}))}, \tag{S7}$$

where $u_{eff}(\boldsymbol{k})\hat{\boldsymbol{u}}(\boldsymbol{k}) = u_R \hat{\boldsymbol{k}} \times \hat{\boldsymbol{z}}$. Note that the matrix $t(\boldsymbol{k})$ is only diagonal when the spin quantization axis is aligned with $\hat{\boldsymbol{u}}(\boldsymbol{k})$. If a different quantization axis is used, the matrix has off-diagonal elements that correspond to the spin-flip amplitudes.

The interface-generated spin current can be separated into two parts,

$$j_\sigma = j_\sigma^I + j_\sigma^{II}, \tag{S8}$$

$$j_\sigma^I = C(\tau^{NM} - \tau^{FM}) \int d\bar{k}_x d\bar{k}_y\, \bar{k}_x T(\boldsymbol{k})_{\sigma c}, \tag{S9}$$

$$j_\sigma^{II} = C\tau^{FM} P \int d\bar{k}_x d\bar{k}_y\, \bar{k}_x T(\boldsymbol{k})_{\sigma\sigma'} m_{\sigma'}, \tag{S10}$$

where $C \equiv eEv_F^3/\hbar(2\pi)^3$. The first part $j_\sigma^I$ is the spin-orbit filtering current that points along $\boldsymbol{f} = \boldsymbol{y}$. The second part $j_\sigma^{II}$ is the spin-orbit precession current that points along $\boldsymbol{m} \times \boldsymbol{y}$. Since the spin-orbit precession current is proportional to the polarization, it vanishes unless one of the layers is ferromagnet.

First, we show that the spin-orbit filtering current points along $\boldsymbol{y}$. Substituting the definition of the scattering tensors, we have:

$$j_\sigma^I = \frac{C}{2}(\tau^{NM} - \tau^{FM}) \int d\bar{k}_x d\bar{k}_y \, \bar{k}_x \text{tr}[t(\mathbf{k})^\dagger \sigma_\sigma t(\mathbf{k})]. \tag{S11}$$

Since $t(\mathbf{k})$ is a diagonal matrix for a spin quantization axis along $\hat{\mathbf{u}}(\mathbf{k})$, we may evaluate the trace in the rotated reference frame $\sigma \in [x', y', z']$ in which $z'$ points along $\hat{\mathbf{u}}(\mathbf{k})$, and then rotate back to the reference frame aligned with the interface ($\sigma \in [x, y, z]$). This gives:

$$j_\sigma^I = \frac{C}{2}(\tau^{NM} - \tau^{FM}) \int d\bar{k}_x d\bar{k}_y \, \bar{k}_x \left(|t^\uparrow(\mathbf{k})|^2 - |t^\downarrow(\mathbf{k})|^2\right) \hat{u}_\sigma(\mathbf{k}). \tag{S12}$$

Substituting the expressions for the transmission amplitudes gives

$$j_\sigma^I = \frac{C}{2}(\tau^{NM} - \tau^{FM}) \int d\bar{k}_x d\bar{k}_y \, \bar{k}_x \left(\frac{\bar{k}_z^2}{\bar{k}_z^2 + u^\uparrow(\mathbf{k})^2} - \frac{\bar{k}_z^2}{\bar{k}_z^2 + u^\downarrow(\mathbf{k})^2}\right) \hat{u}_\sigma(\mathbf{k}), \tag{S13}$$

where for convenience we define $u^{\uparrow/\downarrow}(\mathbf{k}) \equiv u_0 \pm u_{eff}(\mathbf{k})$. Switching to polar coordinates ($\bar{k}_x = r\cos(\phi), \bar{k}_y = r\sin(\phi), \bar{k}_z = \sqrt{1-r^2}$), we may write

$$j_\sigma^I = \frac{C}{2}(\tau^{NM} - \tau^{FM}) \int dr d\phi \, r^2 \cos(\phi) \left(\frac{1-r^2}{1-r^2 + (u_0 + u_R r)^2} - \frac{1-r^2}{1-r^2 + (u_0 - u_R r)^2}\right)$$
$$\times (\delta_{\sigma x}\sin(\phi) - \delta_{\sigma y}\cos(\phi))$$
$$= \frac{C}{2}(\tau^{NM} - \tau^{FM}) \int_0^{2\pi} d\phi \left(\delta_{\sigma x}\cos(\phi)\sin(\phi) - \delta_{\sigma y}\cos^2(\phi)\right) \int_0^1 dr f(r), \tag{S14}$$

where $f(r)$ gives the $r$-dependence of the integrand. Note that $\hat{u}_\sigma(\mathbf{k}) = \delta_{\sigma x}\sin(\phi) - \delta_{\sigma y}\cos(\phi)$. Performing the integral in $\phi$ we arrive at our result,

$$j_\sigma^I = \frac{C}{2}(\tau^{NM} - \tau^{FM})(-\delta_{\sigma y}\pi) \int_0^1 dr f(r). \tag{S15}$$

Computing the integral of $f(r)$ gives the dependence of $j_\sigma^I$ on the scattering parameters $u_0$ and $u_R$, which is not required if only the direction of spin polarization is desired. The final expression for $j_\sigma^I$ is proportional to $\delta_{\sigma y}$, which shows that the spin-orbit filtering current is polarized along **y**.

Second, we show that the spin-orbit precession current points along $\mathbf{m} \times \mathbf{y}$. In polar coordinates we may write $j_\sigma^{II}$ as

$$j_\sigma^{II} = \frac{C}{2}\tau^{FM} P \int dr d\phi \, r^2 \cos(\phi) T(r,\phi)_{\sigma\sigma'} m_{\sigma'}, \tag{S16}$$

where one can show that

$$T(r,\phi)_{\sigma\sigma'} \to S(\phi)\begin{pmatrix} \text{Re}[\bar{t}(r)] & -\text{Im}[\bar{t}(r)] & 0 \\ \text{Im}[\bar{t}(r)] & \text{Re}[\bar{t}(r)] & 0 \\ 0 & 0 & |t^\uparrow(r)|^2 + |t^\downarrow(r)|^2 \end{pmatrix} S(\phi)^\dagger, \quad (S17)$$

where

$$\bar{t}(r) \equiv 2t^\uparrow(r)t^\downarrow(r)^*, \quad (S18)$$

$$S(\phi) \equiv \begin{pmatrix} \cos(\phi) & 0 & \sin(\phi) \\ \sin(\phi) & 0 & -\cos(\phi) \\ 0 & 1 & 0 \end{pmatrix}. \quad (S19)$$

The part of the integral containing $\phi$ can be evaluated

$$\int d\phi \cos(\phi) T(r,\phi)_{\sigma\sigma'} = \pi \text{Im}[\bar{t}(r)] \epsilon_{\sigma\sigma'y} \to \pi \text{Im}[\bar{t}(r)] \begin{pmatrix} 0 & 0 & -1 \\ 0 & 0 & 0 \\ 1 & 0 & 0 \end{pmatrix}, \quad (S20)$$

giving the final result:

$$j_\sigma^{II} = \frac{C}{2} \tau^{FM} P \pi \epsilon_{\sigma\sigma'y} m_{\sigma'} \int dr \, r^2 \text{Im}[\bar{t}(r)]. \quad (S21)$$

The final expression is proportional to $\epsilon_{\sigma\sigma'y} m_{\sigma'} \to \mathbf{m} \times \mathbf{y}$, which shows that the spin-orbit precession current is polarized along $\mathbf{m} \times \mathbf{y}$.

The term 'spin-orbit filtering' arises from the fact that $j_\sigma^I$ is proportional to $|t^\uparrow(\mathbf{k})|^2 - |t^\downarrow(\mathbf{k})|^2$ for each $\mathbf{k}$-vector, so if the transmission probabilities for spins parallel and antiparallel to $\hat{\mathbf{u}}(\mathbf{k})$ differ, a nonvanishing spin current results. This is satisfied when there is interfacial spin-orbit coupling $u_R$ and a spin-independent barrier $u_0$, so that $t^\uparrow(\mathbf{k}) \neq t^\downarrow(\mathbf{k})$. Incident spins may not actually be parallel and antiparallel to $\hat{\mathbf{u}}(\mathbf{k})$, but the result is the same regardless of what quantization axis is chosen. After summing over all $\mathbf{k}$-states, the net spin polarization points along $\mathbf{y}$.

The term 'spin-orbit precession' arises because $j_\sigma^{II}$ is proportional to the tensor $T(\mathbf{k})_{\sigma\sigma'}$, which rotates the vector it is contracted with (in this case $m_{\sigma'} \to \mathbf{m}$) about the spin-orbit field for each $\mathbf{k}$-vector. This can be interpreted as follows: for each $\mathbf{k}$-vector the incoming carriers from the ferromagnetic layer have spins that are parallel (minority carriers) or antiparallel (majority carriers) with the magnetization $\boldsymbol{m}$, and after scattering they each rotate about the $k$-dependent spin-orbit field they see at the interface. After summing over all $k$-states, the net spin polarization points along $\boldsymbol{m} \times \boldsymbol{y}$.

**Note 2. Thermal artefact in the second harmonic Hall voltage measurement**

Figures S2a, b show raw data of the first ($V_{1\omega}$) and second ($V_{2\omega}$) harmonic Hall voltages for the CoFeB/Ti/CoFeB/MgO sample. It is observed that there is an abrupt jump in $V_{2\omega}$ for $B=B_x$, which is due to the anomalous Nernst effect (ANE) originating from the bottom CoFeB with in-plane magnetic anisotropy. The Hall voltage in the *y*-direction is generated by a temperature gradient along the *z*-direction when there is an *x*-component of the magnetization. To verify this, we performed the harmonic measurement for a Ti(2)/CoFeB(4)/Ti(4)/MgO structure, in which the top CoFeB layer with perpendicular magnetic anisotropy is absent. The $V_{2\omega}$ of the sample shows the jump for $B=B_x$, which is identical to that of Fig. S2b. As the $V_{2\omega}$ originating from the ANE effect is irrelevant to the spin-orbit torque, we eliminate this from the raw data when the spin-orbit torque of the sample is analyzed (Fig. 1d of the main text). Figures S3a, b show raw data of the NiFe/Ti/CoFeB/MgO sample, in which a similar thermal voltage in $V_{2\omega}$ is also observed (Figs. S3c, d).

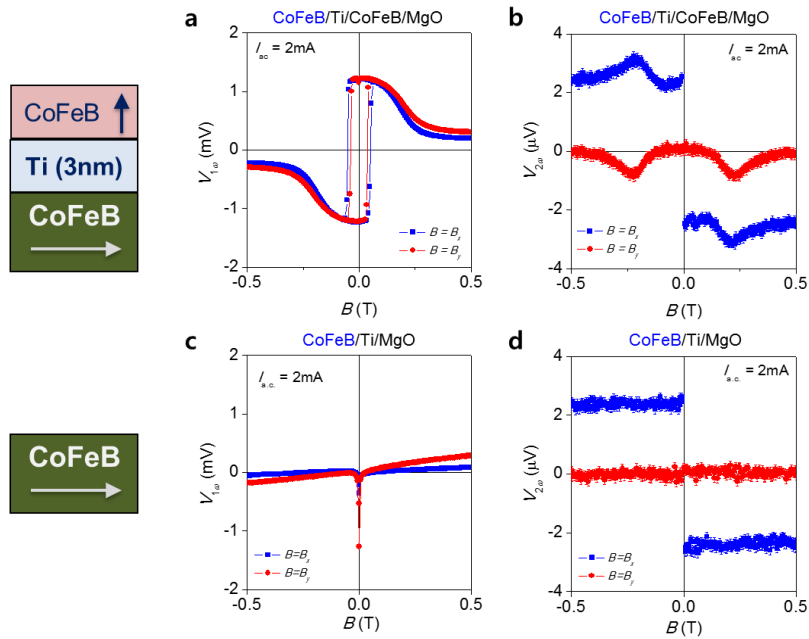

**Figure S2| Raw data of the harmonic measurement for Ti(2 nm)/CoFeB(4 nm)/Ti(3 nm)/CoFeB(1 nm)/MgO and Ti(2 nm)/CoFeB(4 nm)/Ti(4 nm)/MgO samples. a,b,** The first harmonic signal ($V_{1\omega}$) (a) and second harmonic signal ($V_{2\omega}$) (b) for the CoFeB/Ti/CoFeB/MgO structure. **c,d,** $V_{1\omega}$ (c) and $V_{2\omega}$ (d) for the CoFeB/Ti/MgO structure. The measurements are done with an a.c. current of 2 mA.

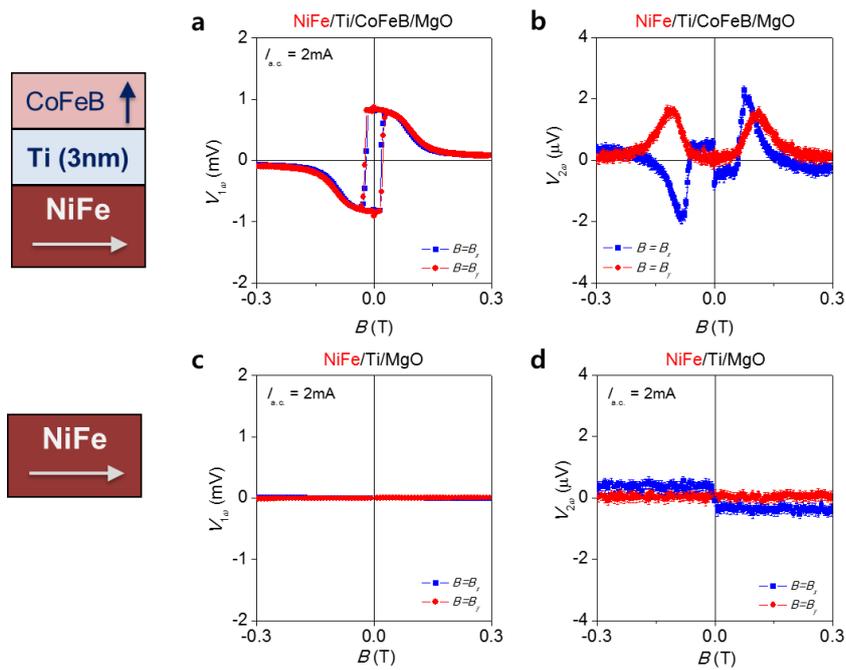

**Figure S3| Raw data of the harmonic measurement for Ti(2 nm)/NiFe(4 nm)/Ti(3 nm)/CoFeB(1.4 nm)/MgO and Ti(2 nm)/NiFe(4 nm)/Ti(4 nm)/MgO samples. a,b,** The first harmonic signal ($V_{1\omega}$) (a) and second harmonic signal ($V_{2\omega}$) (b) for the NiFe/Ti/CoFeB/MgO structure. **c,d,** $V_{1\omega}$ (c) and $V_{2\omega}$ (d) for the NiFe/Ti/MgO structure. The measurements are done with an a.c. current of 2 mA.

**Note 3. Extraction of effective spin Hall angle**

We estimate the effective spin Hall angles of the samples using the relation of $\theta_{SH,eff} = 2eM_s t_F B_D/\hbar|j_e|$ [S8], where $e$ is the electron charge, $M_s$ is the saturation magnetization, $t_F$ is the ferromagnet thickness, $B_D$ is the effective damping-like spin-orbit field, $\hbar$ is the reduced Planck constant, and $j_e$ is the charge current density. $B_D$ of each sample is extracted from the harmonic Hall measurements for a low field regime as shown in Fig. S4 [S9]. We obtain $B_D$ of -22.0±1.0 mT for the Ta sample, -6.5±0.6 mT for the CoFeB/Ti sample, and +2.0±0.2 mT for the NiFe/Ti sample at a current density of $10^8$ A/cm$^2$. We obtain effective spin Hall angles of -0.048±0.002 for the Ta sample, -0.014±0.001 for the CoFeB/Ti sample, and +0.006±0.0006 for the NiFe/Ti sample.

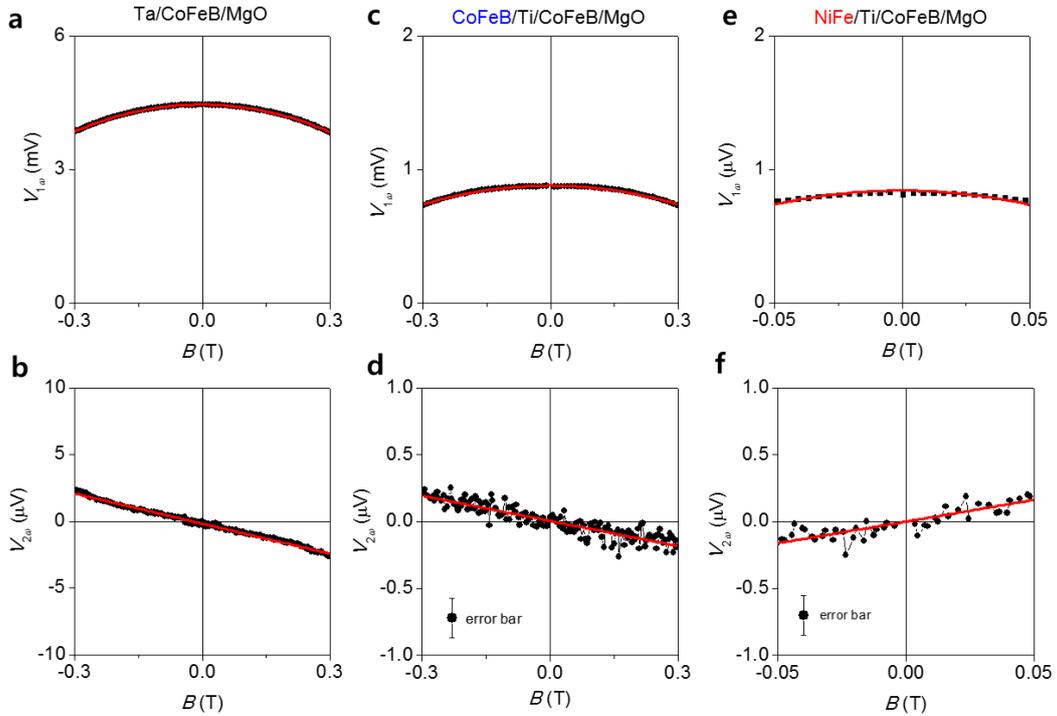

**Figure S4| Estimation of effective damping-like spin-orbit field ($B_D$). a,b,** First and second harmonic signals for Ta/CoFeB/MgO, **c,d,** for CoFeB/Ti/CoFeB/MgO and **e,f,** for Ti/NiFe/Ti/CoFeB/MgO samples. Error bars indicate single standard deviation uncertainties.

**Note 4. Anisotropic magnetoresistance and anomalous Hall resistance of CoFeB and NiFe layers**

We measured anisotropic magnetoresistance (AMR) and anomalous Hall resistance (AHE) of a single ferromagnetic layer of CoFeB (4 nm) and NiFe (4 nm). Note that all samples were covered by a capping layer of MgO(1.6 nm)/Ta(2 nm) to prevent oxidation. AMR is measured by rotating the sample in the film plane with an in-plane magnetic field of 0.3 T. Figure S5a show the AMR of CoFeB and NiFe single layers as a function of the azimuthal angle α, demonstrating that the signs are identical for the CoFeB and NiFe samples. On the other hand, the AHE of the samples measured with out-of-plane field $B_z$ shows opposite sign: positive for CoFeB and negative for NiFe (Fig. S5b). This sign difference in the AHE is consistent with a previous calculation [S10], where Fe and Co show positive anomalous Hall conductivities whereas Ni shows a negative anomalous Hall conductivity.

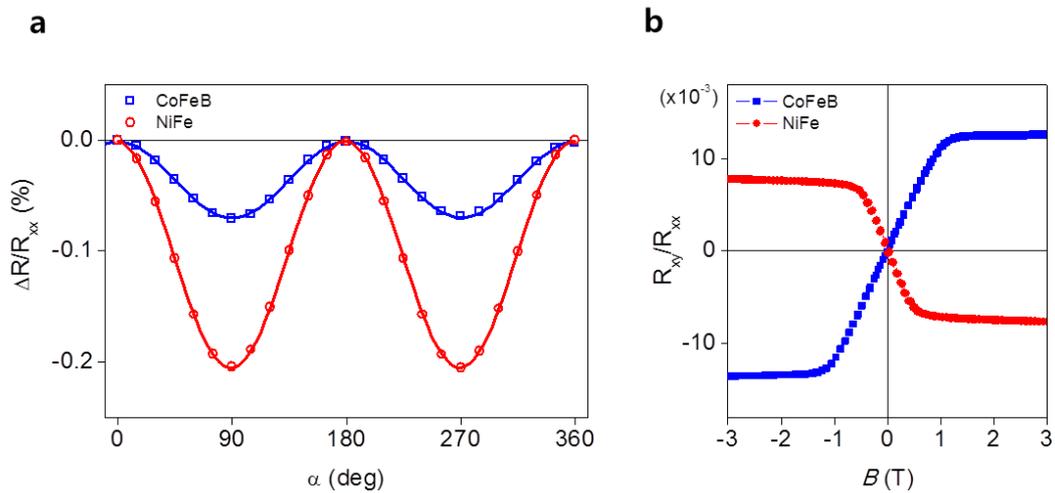

**Figure S5| Anisotropic magnetoresistance (AMR) and anomalous Hall resistance (AHE) of CoFeB(4 nm) and NiFe(4 nm) single layer samples. a,** AMR of CoFeB (blue symbols) and NiFe (red symbols). **b,** AHE of CoFeB (blue symbols) and NiFe (red symbols) structures. α is defined as an angle with respect to the current direction.

**Note 5. Azimuthal angle-dependence of $V_{2\omega}$ for NiFe/Ti/CoFeB/MgO sample**

Figure S6 shows the second harmonic signals ($V_{2\omega}$) measured with in-plane magnetic fields of various azimuthal angles for the NiFe/Ti/CoFeB/MgO sample. This demonstrates a similar angular dependence as the Ta sample and the CoFeB/Ti samples (Figs. 3c,d of the main text), but of the opposite sign.

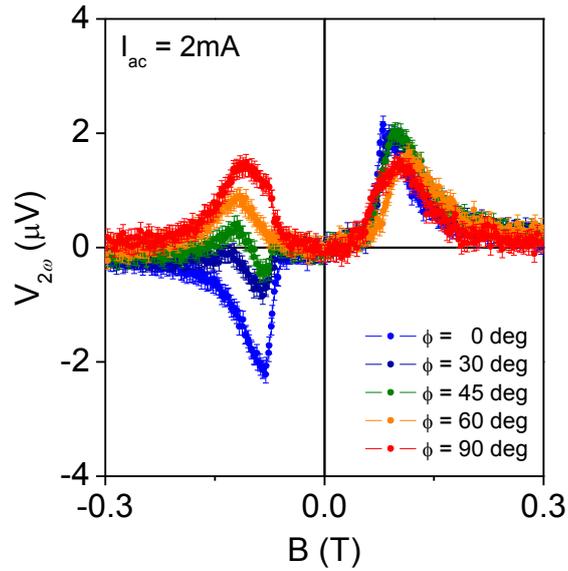

**Figure S6| Azimuthal angle-dependence of $V_{2\omega}$ for Ti(2 nm)/NiFe(4 nm)/Ti(3 nm)/CoFeB(1.4 nm)/MgO sample.** $\phi = 0°$ (90°) is for $B=B_x$ ($B_y$) representing damping (field)-like spin-orbit torque.

**Note 6. The magnetization curves for various dc currents in Ta/CoFeB/MgO structure**

We measured the anomalous Hall signal $R_{xy}$ of the Ta/CoFeB/MgO sample for various d.c. currents in the presence of an in-plane magnetic field ($B_x$) of 10 mT. Figure S7 shows that the hysteresis loop shifts in the positive (negative) $B_z$ direction for negative (positive) d.c. current and the magnitude of the shift increases with the d.c. current. The differences in the centers of the hysteresis loops measured with +$I_{dc}$ and -$I_{dc}$ are plotted in Fig. 3c of the main text. We note that the loop shift is obtained only when $B_x$ is non-zero in Ta/CoFeB/MgO sample.

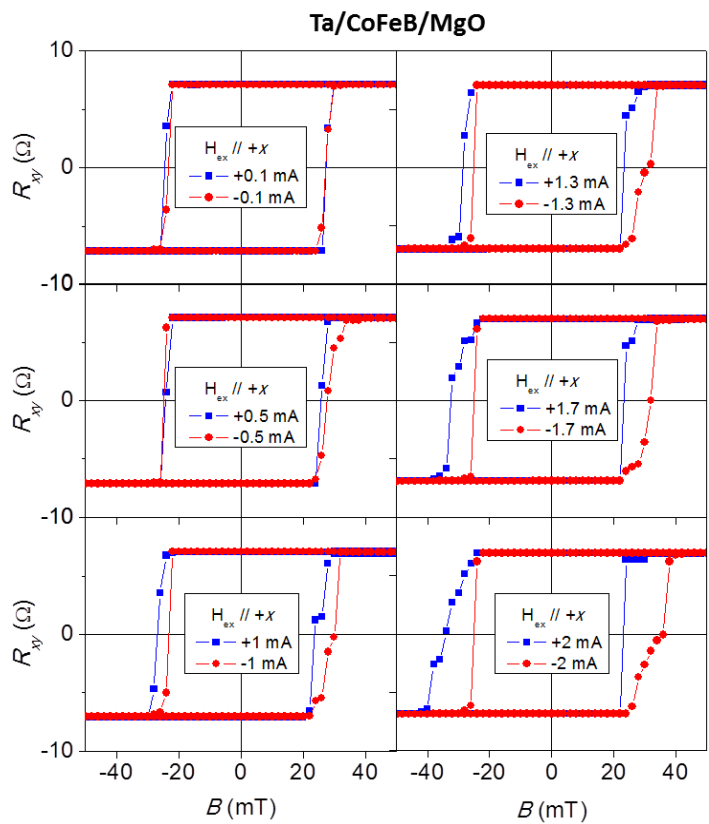

**Figure S7| Magnetization curve versus d.c. current in the presence of $B_x$ for the Ta/CoFeB/MgO sample.** The d.c. current ranges from 0.1 mA to 2 mA and $B_x$ = +10 mT.